\documentstyle[aps,prl,multicol,epsfig]{revtex}

\begin{document}
\title{Ground State Theory of $\delta $--Pu}
\author{S. Y. Savrasov$^{\ast }$ and G. Kotliar$^{+}$}
\author{$^{\ast }$Max-Planck-Institut f\"{u}r Festk\"{o}rperforschung,
Heisenbergstr. 1, \ 70569 Stuttgart, Germany.}
\author{$^{+}$Department of Physics and Astronomy and Center
for Condensed Matter
Theory  \\  Rutgers University, Piscataway,
NJ 08854--8019}
\date{July 1999}
\maketitle

\begin{abstract}
Correlation effects are important for making predictions
in  the  $\delta $ phase of Pu.
Using a  realistic treatment of the intra--atomic Coulomb correlations
we address the long-standing problem of
computing ground state properties.
The equilibrium volume is obtained in good agreement with experiment when taking
into account Hubbard $U$ of the order 4 eV. For this $U,$ the calculation
predicts a 5f$^{5}$ atomic--like configuration with L=5, S=5/2, and J=5/2
and shows a nearly complete compensation between spin and orbital magnetic
moments.
\end{abstract}

\pacs{71.20.-b, 71.27.+a,75.30.-m}

%

Metallic plutonium is a key material in the energy industry and
understanding its physical properties is of fundamental and technological
interest \cite{Pu}. Despite intensive  investigations \cite{PuBook}, 
its extremely rich phase diagram with six
crystal structures as well as its unique magnetic properties are not well
understood.  It is therefore of great interest  to study the ground state
of Pu by modern theoretical methods using first principles electronic
structure calculations, which take into account the possible strong
correlation among the f electrons.

Density functional theory \cite{DFT} in its local density or generalized
gradient approximations (LDA or GGA) is a well-established tool
for dealing with such problems. This theory does an excellent job of
predicting ground-state properties of an enormous class of materials. However,
when applied to Pu \cite {Pucalc1,Pucalc2}, it runs into serious problems.
Calculations of the high-temperature fcc $\delta$ phase
have given an equilibrium atomic volume up to 35\% lower than experiment 
\cite{Pucalc1}. This is the largest discrepancy ever known in density
functional based calculations and points to a fundamental failure of existing
approximations to the exchange-correlation energy functional.

Many physical properties of this phase are puzzling: large
values of the  linear term in the
specific heat coefficient and  of the electrical resistivity
are reminiscent of  the physical properties of strongly-correlated heavy-fermion
systems. On the other hand, the magnetic susceptibility is small and
weakly  temperature
dependent \cite{fournier}. Moreover, early LDA  
calculations \cite{Pucalc2} predicted $\delta $--Pu to be magnetic with a
total moment of 2.1 Bohr magnetons in disagreement with  experiments.
  
The reason for these difficulties has been understood
for a long time:  Pu is located on the
border between light actinides with itinerant 5f--electrons and the heavy
actinides with localized 5f electrons\cite{Johannson} .
Near  this localization-delocalization boundary, the
large intra-atomic Coulomb interaction as well as the itineracy of the f 
electrons
have to be considered on the same footing, and it is expected that  
correlations must be responsible for the anomalous properties. The parameter
governing the importance of correlations in electronic structure
calculations is the ratio between effective Hubbard interaction $U$ and the
bandwidth $W.$ When the  distance between atoms is small, the correlation
effects may be not important, since the hybridization, and consequently
the bandwidth become large. The low-temperature $\alpha $ phase of Pu has an
atomic volume which is 25\% smaller than the volume of $\delta $ phase.
To the extent that the complicated monoclinic structure of the $\alpha $
phase can be modelled by the simplified fcc lattice, it becomes clear that the LDA
or GGA calculations which ignore the large effective $U$ converge to the low
volume 
$\alpha $ phase (for which $U/W < 1$). When volume is increased, this ratio
is turned around, and LDA loses its predictive power. This results in
the long-standing problem of accurate prediction of the volume of  $\delta $--Pu.

In the present work it will be shown that a proper treatment of Coulomb
correlations allows us to compute the equilibrium atomic volume of $\delta $--Pu
in good agreement with experiment. Moreover, our calculations  suggest that
there is a  nearly
complete compensation between the  spin and  the orbital contributions to the total
magnetic moment which is consistent with experiment. Thus the  strong correlation
effects in $\delta $--Pu are  not manifest in the static magnetic properties.

To incorporate the effects of correlations we use the LDA + U approach
of  Anisimov and coworkers \cite{LDA+U}.
This approach recognizes that the failure of LDA is  
related to the fact that it omits 
the Hubbard like   interaction among
electrons in the same shell, irrespectively of  their spin orientation.
A new
orbital--dependent correction  to the LDA functional
was introduced to describe this effect.
In its most recent, rotationally
invariant representation, the correction to the LDA functional has
the following form \cite{LDA+Urecent}: 
\begin{equation}
\Delta E[n]=\frac{1}{2}\sum_{\{\gamma \}}(U_{\gamma _{1}\gamma _{2}\gamma
_{3}\gamma _{4}}-U_{\gamma _{1}\gamma _{2}\gamma _{4}\gamma _{3}})n_{\gamma
_{1}\gamma _{2}}^{c}n_{\gamma _{3}\gamma _{4}}^{c}-E_{dc}  \label{e2}
\end{equation}
where $n_{\gamma _{1}\gamma _{2}}^{c}$ is the occupancy matrix for the
correlated orbital (d or f), and $\gamma $ stands for the combined spin, ($s$),
and azimuthal quantum number,($m$), indexes. The electron--electron
correlation matrix $U_{\gamma _{1}\gamma _{2}\gamma _{3}\gamma
_{4}}=\left\langle m_{1}m_{3}\left| v_{C}\right| m_{2}m_{4}\right\rangle
\delta _{s_{1}s_{2}}\delta _{s_{3}s_{4}}$ can be expressed via Slater
integrals $F^{(i)} $,  $ i=0,2,4,6$ 
in the standard manner
\cite {LDA+Urecent}. The term $E_{dc}$ accounts for the double counting effects.
This  scheme,  known as the ''LDA+U method'',   gives substantial
improvements over the LDA in many cases\cite{LDA+Ureview}. The value of the $U$
matrix
is an input which can  be obtained from a
constrained LDA calculations \cite{CLDA}, or just taken from the experiment.
The philosophy of this approach is that the delocalized s p d electrons
are well described by the LDA while the energetics of the more localized
f electrons require the explicit introduction of the Hubbard U.
In the spirit of this method,  in this work we will treat the s p d electrons
by the generalized gradient approximation (GGA) \cite{GGA} which is believed to
be more accurate that the LDA.

Our implementation of the GGA+U functional is based on the
localized--orbital representation provided by the
linear--muffin--tin--orbital (LMTO)\ method for electronic structure
calculations \cite{OKA}. It is important to include spin--orbit coupling
effects which are not negligible for 5f electrons of Pu. Our calculations include
non-spherical terms of the charge density and potential both within the atomic
spheres and in the interstitial region \cite{Sav}. All low-lying semi-core states
are treated together with the valence states in a common Hamiltonian matrix in
order to avoid unnecessary uncertainties. These calculations are spin polarized
and assume the existence of long--range magnetic order. For simplicity, the
magnetic order is taken to be ferromagnetic \cite{Ferro}.

We now report our results on the calculated equilibrium volume. To analyze
the importance of the correlation effects, our calculations have been
performed for several different values of $U$ varying  from 0 to 4 eV. 
For $U$=4 eV we use standard choice of Slater integrals:
$F^{(2)}$=10 eV, $F^{(4)}$=7 eV, and $F^{(6)}$=5 eV \cite{Pu}. For other U's
we have scaled these values proportionally. 
For each set of $F$'s  a full self--consistent cycle minimizing the
LDA/GGA+U functionals has been performed for a number of atomic volumes. We
calculated  the total energy $E$ as a function of both $V$
and $U$. For fixed $U,$ the theoretical equilibrium, $V_{calc},$ is given by
the minimum of $E(V)$. Fig. 1 shows the dependence of the
calculated--to--experimental equilibrium volume ratio $V_{calc}/V_{exp}$ as
a function of the input $U.$ It is clearly seen that the $U$=0 result (LDA)
predicts an  equilibrium volume which is 38\%  off the
experimental result  and the use
of GGA gives only slightly improved result ($V_{calc}/V_{exp}$=0.66). On the
other hand, switching on a very large repulsion between 5f electrons
obviously leads to  an overestimate of the inter-atomic distances. An optimal $U$
deduced from this analysis is found to be close to 4 eV when using the GGA
expressions for  the exchange and correlation.

This estimate of the intra-atomic correlation energy is in excellent agreement
with the published conventional data \cite{PuU}: The value of $U$ deduced
from the total energy differences was found to be 4.5 eV. Atomic spectral
data give similar value close to 4 eV. Thus, it is demonstrated how
significant it is to properly treat Coulomb correlations in predicting the
equilibrium properties of this actinide.

We now discuss the calculated GGA+U electronic structure of $\delta $--Pu
for the optimal value of $U$=4 eV. Fig. 2 shows the energy bands in the vicinity
of the Fermi level. They originate from the extremely wide 6s--band strongly
mixed with the 5d--orbitals which are 
strongly hybridized among themselves.
The resulting band complex has a  bandwidth
of the order  of 20 eV.
On top of this structure there exist a weakly hybridized set of
levels originating from the 5f--orbitals.

In order to understand the physics behind the formation of spin and orbital
moment in the f--shell, it is instructive to visualize the orbital
characters as ''fat bands''\cite{Fat}. 
The one--electron wave function has an expansion 
$\psi _{{\bf k}j}({\bf r})=\sum A _{lms}^{{\bf k}j} \phi _{lms}({\bf r})$
where $\phi _{lms}({\bf r})$
are the solutions of the radial Schr\"odinger equation
normalized to unity within atomic sphere.
The  information about partial $lms$ characters
of the state with given ${{\bf k}j}$ is contained in the coefficients 
$|A _{lms}^{{\bf k}j}|^2$.
Sum over all $lms$ in the latter quantity gives unity (we neglect by a small contribution
from the interstitial region) since one band carries one electron per cell. 
At the same time, sum over all $j$ in $|A _{lms}^{{\bf k}j}|^2$ is also equal
to one since each $lms$ describes one state. 
Fixing a particular $lms$, we can visualize this partial character
on top of the band structure by widening each band $E_{{\bf k}j}$ proportionally
to $|A_{lms}^{{\bf k}j}|^2$. A maximum width $\Delta $ which corresponds to 
$\sum _j |A_{lms}^{{\bf k}j}|^2$=1 should be 
appropriately chosen. Now, at the absence of hybridization, each band 
originates from a particular $lms$ state, and therefore there exists only one
''fat band'' for given $lms$ which has the maximum width $\Delta $. When hybridization
is switched on, there can be many bands which have the particular $lms$ character,
they will all be widened as $\Delta |A_{lms}^{{\bf k}j}|^2$, while the 
sum of individual widths for all bands is now equal to $\Delta $. 
The width of  the band is then proportional to its $lms$ character.
This technique \cite{Fat} gives us an important 
information on the distribution of atomic levels as
well as their hybridization in a solid. 
For f--electrons of Pu, it is
convenient to work in the spherical harmonics representation in which the
f--f block of the Hamiltonian is found to be nearly diagonal.

The result of such ''fat bands'' analysis for 5f--orbitals is shown on Fig.
2. In order to distinguish the states with different $m$'s and spins we have
used different colors. (-3 $\equiv $ red, -2 $\equiv $ green, -1 $\equiv $
blue, 0 $\equiv $ magenta, +1 $\equiv $ cyan, +2 $\equiv $ yellow, +3 $%
\equiv $ gray). Two consequences are seen from this coloured spaghetti:
First, spin--up and spin--down bands are all split by the values governed
by the effective $U$ and the occupancies of the levels. These are just the
well known lower and upper Hubbard subbands. Second, only spin--up states
with $m$=-3,-2,-1,0, and +1 are occupied while all other states are empty.
This simply implies 5f$^{5}$ like atomic configuration for $\delta $--Pu
which is filled according to the Hund rule. Note that
spin-orbit coupling is crucial for the existence of such an occupation scheme.
In the absence of spin orbit coupling 
the occupancies of the levels with $\pm m$ are the same which
automatically produces zero orbital moment.

Besides  providing  the experimentally observed  volume of
the  $\delta $--Pu, 
our calculation   suggests a simple picture of the electronic
structure of this material   and sheds new light on 
its  puzzling physical properties  discussed in the introduction.

The  ''fat bands'' shown in Fig. 2, suggest a physical
picture in which   the f electrons
are  in    atomic states   
forming a  multiplet of the  5f$^5$ configuration with L=5, S=5/2
spin orbit coupled to  J=5/2. Crystal fields can split this multiplet
into a 
doubly degenerate state transforming according to $\Gamma_7$ 
representation of the cubic group and a  quartet   
transforming according to $\Gamma_8$ representation \cite{PuBook}, but
cannot remove the orbital  degeneracy completely. 
In a dynamic picture,  the f electrons will fluctuate between the
degenerate configurations, until this
degeneracy  is removed by the 
Kondo effect with the delocalized electrons in the s-p-d band.
Therefore 
the experimentally observed 
characteristic  heavy fermion  behavior in this system,
namely,  the  large   high--temperature resistivity and
the  large   linear  T coefficient of the specific
heat  arises naturally in this picture
\cite{PuBook}.

This heavy fermion behavior however should  not appear
in the magnetic susceptibility.
The     GGA + U calculation suggests that
the magnetic moment of the low lying configurations
of the f electrons  is much smaller than the $5 \mu_B $  that one
would  obtain if  we ignore  
the orbital angular momentum and assumed that the    spin is fully polarized.
The  combination of
strong Coulomb interactions and spin--orbit coupling 
reduce the crystal--field effects  and give rise to a large
magnetic moment which nearly cancels the spin moment. 
In an atomic picture, 
the $5f^5$ configuration with L=5, S=5/2 and
J=5/2 has a   total moment   given by $M_{tot}$ $=\mu _{B}gJ=0.7\mu
_{B}$, with Lande's g-factor of  0.28.
This simple relation breaks down in the presence
of crystal fields, but in both  the $\Gamma_7$
or the $\Gamma_8$ representation  the g factor is  further
reduced from the  atomic estimate.

The GGA + U calculation  gives a  spin moment
$ M_{S}$ 5.1 Bohr magnetons which is slightly increased relative to 
the  5 Bohr magnetons expected in a pure f$^{5}$ atomic configuration
due to the polarization of the  band electrons outside the muffin--tin shell.
Evaluation of the
orbital and total moments is in general a more difficult problem \cite{Cal}.
We have  estimated   the average of $\langle {\bf k}j\left|
l_{z}\right| {\bf k}j\rangle $ summed over all occupied states $|{\bf k}%
j\rangle .$ This leads to  a value  $M_{L}$=-3.9 $\mu
_{B.}$ for the orbital moment.
 The total calculated moment $M_{tot}=M_{S}+M_{L}$ is thus reduced to
1.2 $\mu _{B}$ . It worth  noting  that an atomic analogue of this estimate,
$ M_{tot}$ $=\mu _{B}\left| L-2S\right| $ gives  exactly zero for our  5f$^{5}$ ground
state\cite{Another}. A remarkable outcome   of the  calculation is clearly
seen: {\em A nearly complete compensation of spin and orbital contributions
occurs} for metallic $\delta $--Pu. 

In this picture  the   weakly  temperature independent susceptibility which 
is observed in $\delta $--Pu  
\cite{PuBook,fournier} is the result of a 
a very large Van Vleck contribution and of a very small magnetic moment
which results from the near cancellation of two large orbital and spin moments.

In a recent paper \cite{Eriksson} Eriksson and
coworkers introduced a different approach to the
anomalous properties of $\delta$ plutonium. In their calculation 
a fraction of  the f-electrons is treated as core electrons
while the rest are treated as delocalized. 
Using  a combination of the
constrained LDA calculation with the atomic multiplets data
they obtain  the  correct equilibrium volume
when four f-electrons  are part of the core, while one f electron
is itinerant.
The basic difference between the methods, is the different
treatment of the f electrons. In this paper 
all the f electrons on equal footing,
and their itineracy 
is reduced by the Hubbard U
relative to the predictions of  LDA or GGA calculations.
Since our approach and that of Eriksson et. al. lead
to different ground state configurations of the localized
f electrons ($f^5$ vs $f^4$), further experimental  spectroscopic
studies of $\delta $--Pu  would be of interest.

In conclusion, using a realistic value of the Hubbard $U$=4 eV incorporated
into the density functional GGA calculation, we have been able to
describe ground state properties of $\delta $--Pu in good agreement with
experimental data. This theory correctly predicts the equilibrium volume of
the $\delta $ phase and suggest that nearly complete cancellation occurs
between spin and orbital moments. The main shortcoming of the  present calculation
is the assumed long range spin and orbital order. 
This is the essential limitation of the LDA+U approach (or of any
{\it static} mean field theory )  in order   to capture the
effects of correlations this approach it  has to impose some form of long--range order.
Static mean field theories are  unable to capture subtle many--body effects such 
as the formation of local moments and
their subsequent quenching via the Kondo effect.
These deficiencies will be removed by  ab initio 
{\it dynamical} mean field \cite{DMFT} calculations for which codes
are  currently being developed. 
We believe however, that our main conclusions, i.e. that  
correlations  lead to the correct lattice constant and a reduction of the moment,
relative to the LDA results, are robust consequences                   
of the strong correlations presented in this material, and will be reproduced
by more accurate treatments of the electron correlations.

The authors are indebted to E. Abrahams, O. K. Andersen, O. Gunnarsson, A. I.
Lichtenstein, and J. R. Schrieffer for many helpful discussions. The work
was supported by the DOE division of basic energy sciences, grant
No. DE-FG02-99ER45761.

FIGURE CAPTIONS

Fig.1. Calculated theoretical volume (normalized to the experiment) 
of $\delta $--Pu as a function of the Hubbard $U$ within the LDA+U
and GGA+U approaches. 

Fig.2. Calculated energy bands of $\delta $--Pu using GGA+U method
with $U$=4 eV. Spin and orbital characters of the f-bands are shown 
with the color:
($m$=-3 $\equiv $ red, -2 $\equiv $ green, -1 $\equiv $
blue, 0 $\equiv $ magenta, +1 $\equiv $ cyan, +2 $\equiv $ yellow, +3 $%
\equiv $ gray). Boxes from the left and from the right show approximate
positions of the f levels.

%


\begin{references}
\bibitem{Ferro}  In fact, our own and previous \cite{Pucalc2} total energy
calculations suggest that the antiferromagnetic state has slightly lower
energy.  A realistic theory should not assume any
long--range order at all. Unfortunately, it is very hard to treat
short--range order with the existing techniques (through the use of
supercells), while alternative methods based on dynamical mean--field theory
and Green functions are still in developmental stage.
\bibitem{Another}  Note that we have found another metastable solution of
the functional \ref{e2} corresponding to the occupation of the levels with $%
m=-3,-2,-1,+1,+2.$ The orbital moment of this configuration is 
about 3 and carries a large magnetic moment.
The energy of this configuration is however found to lie 0.5 Ry higher than
the energy of the ground state configuration with high orbital moment.
\end{references}
\end{document}